\documentclass[%
 reprint,
nofootinbib,
 amsmath,amssymb,
 aps,
 prd,
]{revtex4-2}

\usepackage{graphicx}% Include figure files
\usepackage{dcolumn}% Align table columns on decimal point
\usepackage{bm}% bold math
\usepackage{color}
\usepackage[dvipsnames]{xcolor}
\usepackage{lineno}
\usepackage{booktabs}
\usepackage{bbding}
\usepackage{tabularx}
\usepackage{subcaption}
\usepackage{multirow}
\usepackage{array}
\usepackage{float}
\usepackage{hyperref}
\usepackage{amsmath}
\usepackage[normalem]{ulem}
\usepackage{comment}

\begin{document}

\preprint{APS/123-QED}

\def\ef#1{\textcolor{blue}{[\textbf{EF}: #1]}}
\def\cdl#1{\textcolor{red}{[\textbf{CDL}: #1]}}
\def\snc#1{\textcolor{red}{\tt[SN: #1]}}
\def\snt#1{\textcolor{red}{#1}}

\def\mmc#1{\textcolor{pink}{\tt[MM: #1]}}
\newcommand{\apjl}{Astrophys. J. Lett.}
\newcommand{\mnras}{Mon. Not. Roy. Astron. Soc.}
\newcommand{\jcap}{JCAP}
\newcommand{\aap}{Astron. Astrophys.}

\title{\boldmath Exploring Hu-Sawicki-like modified gravity with Genetic Algorithms}
%\mmc{"Genetic Algorithms selection of $f(R)$ modifications of gravity"?} \\ or 

\author{Chiara De Leo}
 \email{chiara.deleo@uniroma1.it}
\affiliation{%
 \itshape Physics Department, Sapienza University of Rome, P.le A. Moro 5, 00185 Roma, Italy,\\
 Istituto Nazionale di Fisica Nucleare (INFN), Sezione di Roma, P.le A. Moro 5, I-00185, Roma, Italy, \\
 \itshape Physics Department, Tor Vergata University of Rome, Via della Ricerca Scientifica 1, Roma, 00133, Italy,\\
 \itshape INAF - Osservatorio Astronomico di Roma, via Frascati 33, 00078 Monte Porzio Catone, Italy
}%

\author{Elisa Fazzari}
 \email{elisa.fazzari@uniroma1.it}
\affiliation{%
 \itshape Physics Department, Sapienza University of Rome, P.le A. Moro 5, 00185 Roma, Italy,\\
 Istituto Nazionale di Fisica Nucleare (INFN), Sezione di Roma, P.le A. Moro 5, I-00185, Roma, Italy, \\
 \itshape Physics Department, Tor Vergata University of Rome, Via della Ricerca Scientifica 1, Roma, 00133, Italy
}%

\author{Matteo Martinelli}
 \email{matteo.martinelli@inaf.it}
\affiliation{%
 \itshape INAF - Osservatorio Astronomico di Roma, via Frascati 33, 00078 Monte Porzio Catone, Italy\\
 Istituto Nazionale di Fisica Nucleare (INFN), Sezione di Roma, P.le A. Moro 5, I-00185, Roma, Italy
}%

\author{Savvas Nesseris}
 \email{savvas.nesseris@csic.es}
\affiliation{%
\itshape Instituto de Física Téorica, UAM-CSIC, Universidad Autónoma de Madrid, Cantoblanco, 28049 Madrid, Spain
}%

\date{\today}% It is always \today, today,

\begin{abstract}
We investigate whether viable Hu-Sawicki-like $f(R)$ models can produce deviations from $\Lambda\mathrm{CDM}$ that can be tested against current background cosmological data. We adopt a machine-learning approach based on Genetic Algorithms (GA) to reconstruct analytical perturbations around the Hu-Sawicki class of models. We develop a pipeline that interfaces the \texttt{GATO} GA code with the \texttt{CANDI} cosmology code. Each $f(R)$ function generated by the GA is first tested against theoretical viability conditions, including stability, the recovery of a standard matter-dominated epoch, the General Relativity limit, and chameleon screening mechanism. Viable candidates are then passed to \texttt{CANDI} to reconstruct the corresponding background cosmology and are tested against DESI DR2 BAO measurements and the Pantheon+ Type Ia supernova catalogue.
%\newline
The deviations we find are largest at late times, where the lower curvature makes modified-gravity effects more relevant, and are rapidly suppressed at higher redshift, in agreement with the imposed matching to the matter-dominated era. To further quantify deviations from the standard cosmological model, we compute the $Om(z)$ diagnostic. It shows only a very small departure from the constant $\Lambda\mathrm{CDM}$ behaviour. The effective dark energy equation of state associated with the reconstructed $f(R)$ function also evolves only weakly, showing a mild transition from an effective quintessence-like nature to an effective phantom-like regime.
Overall, our results indicate that, within perturbations around the Hu-Sawicki class of models, current background data allow only limited deviations from $\Lambda\mathrm{CDM}$.
\end{abstract}

\maketitle

%\tableofcontents

\section{Introduction}
%\ef{List of acronyms: GR, EH, DE, FLRW, EoS} \\
Understanding the physical mechanism behind the accelerated expansion of the Universe remains one of the major challenges of modern cosmology. In this respect, the recent Baryon Acoustic Oscillation (BAO) results from the Dark Energy Spectroscopic Instrument (DESI) \citep{DESI:2024mwx,DESI:2024uvr,DESI:2024kob,DESI:2024lzq,DESI:2024aqx,DESI:2025zgx,DESI:2025zpo,DESI:2025qqy} have played a crucial role in reassessing the standard cosmological scenario, i.e. $\Lambda$ Cold Dark Matter ($\Lambda \mathrm{CDM}$) model. Current data appear to favor the possibility of an evolving dark energy (DE) component, i.e. a DE equation of state (EoS) parameter $w(a)$ evolving with the expansion, rather than a strictly constant ($\Lambda$) term. In the commonly adopted Chevallier-Polarski-Linder (CPL) parametrization~\citep{Chevallier:2000qy,Linder:2002et}, i.e. $w(a)=w_0+w_a(1-a)$, this preference is typically associated with a present-day quintessence-like EoS ($w_0>-1$), together with a negative evolution parameter ($w_a<0$), implying a transition towards the phantom regime at earlier times.

The robustness of this indication has been widely investigated after the DESI releases, both with respect to the assumed DE parametrization and to the specific dataset combinations adopted. In particular, several analyses have shown that the preference for a dynamical DE scenario seems to be not merely an artifact of the CPL form, but persists when alternative parameterizations of DE are considered~\citep{DESI:2025fii, Giare:2024gpk, Wolf:2025jlc}. At the same time, it has also been argued in ~\citep{Nesseris:2025lke} that the inclusion of additional degrees of freedom in the EoS, subsequently marginalized over, reduce the statistical evidence for an evolving DE to a negligible level. This picture is further supported by model-independent cosmographic analyses~\citep{Fazzari:2025lzd} and by non-parametric reconstructions of the DE EoS or energy density~\citep{Goh:2024exx, Ormondroyd:2025exu, Ormondroyd:2025iaf, Berti:2025phi, DESI:2025fii, Gonzalez-Fuentes:2025lei, Mukherjee:2025ytj}.

In this framework, the absence of a fundamental physical explanation for an evolving DE component has motivated the development of a wide variety of alternative cosmological scenarios. Among them, modified gravity models represent an important alternative to scalar-field DE, since they aim to explain the accelerated expansion of the Universe through a modification of the gravitational sector itself \citep{Sotiriou:2008fR}.
Particular attention over the last decades has been devoted to $f(R)$ gravity theories~\citep{Nojiri:2006ri,Sotiriou:2008fR, Faraoni:2010beyond}. In these models, the Einstein-Hilbert action is generalized by replacing the Ricci scalar $R$ with a generic function $f(R)$, extending the Einstein field equations in General Relativity (GR). The first cosmological application was proposed by Starobinsky \citep{Starobinsky:1980te}, who introduced an $R^2$ correction to drive inflation. Since then, $f(R)$ models have been widely investigated from several perspectives, and many different functional forms have been proposed in the literature \citep{Sotiriou:2008fR, Faraoni:2010beyond, Capozziello:2011et}.

Among late-time $f(R)$ models, some of the most promising and widely studied examples are the Hu-Sawicki (HS) \citep{Hu:2007nk}, Starobinsky \citep{Starobinsky:2007hu}, Tsujikawa \citep{Tsujikawa:2007xu}, and Appleby-Battye \citep{Appleby:2009uf} models. These are particularly interesting because they can satisfy the main theoretical and observational viability requirements, including stability conditions, a positive scalaron mass squared avoiding the presence of tachyonic particles, a standard matter-dominated epoch, the recovery of the GR limit, and the activation of a chameleon screening mechanism guaranteeing the recovery of GR limit in high-density environments and while preserving an effect on cosmological scales \citep{Sotiriou:2008fR, Amendola:2006we, Brax:2008hh}.
The crucial point of interest is that the modified geometric part of the field equations can be interpreted as an effective DE component with an evolving EoS parameter $w(a)$ \citep{Hu:2007nk, Amendola:2006we, Martinelli_2009}.

In principle, an $f(R)$ cosmology can be tested in two different ways. The standard approach consists in assuming a specific functional form for $f(R)$, characterized by a set of free parameters, and then constraining these parameters with cosmological data. However, this procedure inevitably introduces a certain degree of model dependence, since the final results can be affected by the chosen parametrization. In this work, we adopt instead a machine learning approach based on stochastic optimization with symbolic regression, namely the Genetic Algorithms (GA), on $f(R)$ with the aim of exploring perturbations around the HS scenario. 

In particular, we specifically pick the HS model as it satisfies all theoretical checks and has an effective DE EoS parameter $w_\mathrm{eff}$ curve of dynamical DE which crosses -1, as shown in \citep{Hu:2007nk}, but also is one of the most well tested $f(R)$ models in the literature with stringest constraints expected with forthcoming data \cite{Euclid:2023tqw}. Thus, in this scenario we can test if the HS $f(R)$ is really able to reproduce a dynamical DE scenario with background data when considering DESI DR2 BAO with Type Ia supernovae (SNe) data.

A main advantage of using the GA in conjunction with the HS scenario is that solving the full set of differential equations for the $f(R)$ models is particularly computationally expensive, thus making the exploration of the full functional space very computational expensive. Thus, exploring with the GA around a particular model we avoid this issue while still maintaining a lot of freedom in the analysis \cite{Kamerkar:2022dfu}. Moreover, the GA are much more efficient at this kind of explorations than manually exploring the variants of Hu-Sawicki-like models \cite{Perez-Romero:2017njc}.

Second, given the nature of the reconstruction, a random $f(R)$ function in general will not be viable, in the sense of having a radiation, matter, and acceleration epochs, due to the plethora of the different stability points of the corresponding autonomous system \citep{Amendola:2006we}. Thus, our analysis is in a sense a proof of concept and is similar in spirit to that of \citet{Kamerkar:2022dfu}, where the GA was used to explore small perturbations around slow rolling inflationary models.

The GAs have already been applied in several cosmological contexts, e.g. see Refs. \citep{Bogdanos:2009ib, Nesseris:2010ep, Nesseris:2012tt, Nesseris:2013bia, Sapone:2014nna, Arjona:2020kco, Arjona:2020doi, Kamerkar:2022dfu, Goh:2024exx}, also in forecasts for forthcoming large scale structure surveys \citep{Euclid:2020ojp, Euclid:2021cfn, Euclid:2021frk}, but also in other areas such as particle physics \citep{Abel:2018ekz, Allanach:2004my, Akrami:2009hp} and astronomy and astrophysics \citep{Rajpaul_GA, Ho:2019zap}. Here, we interface the GA reconstruction code with a dedicated pipeline in order to find the resulting $f(R)$ functions that best fit current background cosmological data.

The paper is structured as follows. In Sec.~\ref{sec:fR_1}, we review the main theoretical aspects of $f(R)$ gravity and discuss the viability conditions relevant for late-time cosmology. In Sec.~\ref{sec:Genetic_algorithms}, we describe the Genetic Algorithm approach and its implementation for the reconstruction of $f(R)$ functions. In Sec.~\ref{sec:dataset_methodology}, we present the observational datasets and the pipeline used in the analysis. Finally, in Sec.~\ref{sec:results}, we summarize our main results within the best-fit reconstructed $f(R)$ function.

\section{$f(R)$ modified gravity}\label{sec:fR_1}
We study deviations from GR within the framework of $f(R)$ modified gravity, where the Einstein-Hilbert (EH) action, linear in the Ricci scalar $R$, is generalized by replacing $R$ with a generic scalar function $f(R)$ \cite{Sotiriou:2008fR, DeFelice:2010aj}. The total action is therefore given by

\begin{equation}
    S[g_{\mu\nu}]=\frac{1}{2\kappa^2}\int\mathrm{d}^{4}x\sqrt{-g} f(R) + S_\mathrm{M}[g_{\mu\nu},\psi] \, ,
    \label{eq:azione_fR}
\end{equation}
where, adopting the metric signature $(-,+,+,+)$, $\kappa^2 \equiv 8 \pi G_\mathrm{N}$ is the Einstein constant (in natural units), $G_\mathrm{N}$ is the Newtonian gravitational constant, $g$ is the determinant of the metric tensor $g_{\mu\nu}$, $S_\mathrm{M}$ is the matter action, and $\psi$ stands for matter fields. 

Varying the action in Eq.~\eqref{eq:azione_fR} with respect to the metric $g^{\mu\nu}$ one obtains the generalized field equations\footnote{In the metric formalism, the metric tensor is the only gravitational variable varied in the action, while the connection is assumed to be the Levi-Civita connection associated with $g_{\mu\nu}$.}
\begin{equation}
    f_R R_{\mu\nu}- \frac{1}{2}g_{\mu\nu}f(R)+ g_{\mu\nu}\Box f_R- \nabla_{\mu}\nabla_{\nu}f_R=\kappa^2\, T_{\mu\nu} \;,
    \label{eq:fR_field_equations}
\end{equation}
where $f_R\equiv df/dR$ is the derivative of $f(R)$ with respect to the Ricci scalar, $T_{\mu\nu}$ is the stress-energy tensor, and $\Box$ is the usual d'Alembert operator.

It is useful to parametrize the deviation from the EH action by introducing a function $F(R)$ such that 
\begin{equation}
    f(R)=R+F(R), \qquad \mathrm{with}  \qquad F_R\equiv dF/dR \,.
    \label{eq:fR_FR}
\end{equation}
This notation makes explicit the departure from the GR limit, which is recovered when the additional contribution $F(R)$ becomes negligible or a constant term.

Considering a spatially flat, homogeneous, and isotropic universe described at large scales by the Friedmann – Lema\^itre – Robertson – Walker (FLRW) line element \cite{weinberg}, i.e.

\begin{equation}
ds^{2}=-dt^{2}+a^{2}(t) \left(dx^2+ dy^2+ dz^2\right) \;,
\label{eq:FLRW_lineele}
\end{equation}
where $a(t)$ represents the scale factor, and $t$ denotes synchronous time, one can derive the modified Friedmann equation \cite{Sotiriou:2008fR}

\begin{equation}
    H^2-F_R(H H^{\prime}+H^2)+\frac{F}{6}+H^2 F_{RR} R^{\prime}=\frac{\kappa^2}{3}\rho \;,
    \label{eq:Friedmann_mod}
\end{equation}
where $^{\prime}\equiv d/d \ln a$. This equation follows from the $00$ component of Eq.~\eqref{eq:fR_field_equations}, together with the stress-energy tensor of a perfect fluid i.e.
$T_{\mu\nu}^{\text{p.f.}}=(\rho+P)u_{\mu}u_{\nu}+P g_{\mu\nu}$, where $\rho$ is the energy density and $p$ is the pressure, linked by the the EoS $p=w \rho$ with $w$ being the EoS parameter. In order to remain close to a late-time $\Lambda\mathrm{CDM}$ scenario, we adopt as source terms matter ($w=0$) and we will not directly take into account a DE source or a cosmological constant ($w=-1$), instead we will consider $f(R)$ expressions that contain the cosmological constant scenario as a limiting case. 

The modified Friedmann equation, Eq.\eqref{eq:Friedmann_mod}, reduces to the standard scenario when considering $F=-2\Lambda$, yielding to 
\begin{equation}
    H^2=\frac{\kappa^2}{3}\left(\rho+\rho_{\Lambda}\right)\;,
    \label{eq:Friedmann_LambdaCDM}
\end{equation}
where $\rho_{\Lambda}\equiv \Lambda/\kappa^2$ is the cosmological constant term. Explicating the energy-density components for a late-time Universe where radiation and neutrinos are subdominant, this expression can be recast as
\begin{equation}
    H^{2}(a)=H_0^{2} \left(\Omega_\mathrm{m,0}\,a^{-3}+1-\Omega_\mathrm{m,0} \right) \;, 
    \label{eq:Friedmann_LambdaCDM_norm}
\end{equation}
where we have introduced the dimensionless density parameters
\begin{equation}
\Omega_i \equiv \frac{\rho_i}{\rho_{\mathrm{cr},0}} 
\qquad \mathrm{with} \qquad \rho_{\mathrm{cr},0} =\frac{3H_0^2}{\kappa^2} \,,
\end{equation}
with $H_0$ the Hubble constant and $\Omega_{\Lambda}=1-\Omega_\mathrm{m,0}$ in a spatially flat Universe at late times, where $0$ denotes the today-values hereafter.

In general, for a specific $f(R)$ function the modified Friedmann equation must be solved together with the expression of the Ricci scalar for the FLRW metric,
\begin{equation}
    R=12 H^2 + 6 H H^{\prime} \,.
    \label{eq:R_FLRW}
\end{equation}

Defining the auxiliary variables $y_H = (H^2/m^2)-a^{-3}$ and $y_R = (R/m^2)-3a^{-3}$ with $m^2=\kappa^2 \rho /3$, the Friedmann equation together with the Ricci scalar function can be rewritten as a system of two ordinary differential equations \cite{Hu:2007nk}, i.e.

\begin{subequations}\label{eq:y_ODE}
\begin{align}
y_H^{\prime}&=\frac{y_R}{3}-4y_H \, ,
\label{eq:yH_prime}
\\
y_R^{\prime}&=9a^{-3}-\frac{1}{y_H+a^{-3}}\frac{1}{m^2F_{RR}}\biggl[y_H-    F_R\biggl(\frac{y_R}{6}-y_H\nonumber\\&\qquad\qquad\qquad\qquad-        \frac{a^{-3}}{2}\biggr)\frac{F}{6m^2}\biggr] .
\label{eq:yR_prime}
\end{align}
\end{subequations}
To complete and solve the system, we consider the initial conditions at high redshifts from the $\Lambda \mathrm{CDM}$ limit, using Eq.~\eqref{eq:Friedmann_LambdaCDM_norm}, that means imposing that $\lim_{R\to\infty} F(R) = -2 \Lambda $.

The modified cosmological dynamics encoded in the $f(R)$ function can be recast as an effective DE component   defining the effective EoS parameter as \cite{Hu:2007nk, Amendola:2006we}
\begin{equation}
w_\mathrm{eff}=-1-\frac{1}{3}\frac{y^{\prime}_H}{y_H} \;.
\label{eq:wDE_yH}
\end{equation}
It is clear that this quantity should be understood as an effective description of the modified gravitational dynamics rather than as the EoS of a fundamental fluid, but it is a useful tool that allows us to compare the background expansion generated by a given $f(R)$ model with phenomenological DE parameterizations \cite{Amendola:2006we}.

Finally, the metric $f(R)$ formalism can be recast as a scalar-tensor theory that in general is useful to reduce the field equation to second order from fourth order. In this work, this representation will be used for formulating the theoretical conditions that an $f(R)$ model should satisfy in order to avoid instabilities and remain compatible with local and cosmological constraints.

The action in Eq.~\eqref{eq:azione_fR} can be rewritten in its dynamically equivalent formulation in the so-called Jordan frame \cite{Sotiriou:2008fR}, taking the form:

\begin{equation}
S_{J}[g_{\mu\nu}]=\frac{1}{2\kappa^2}\int\mathrm{d}^{4}x\sqrt{-g}\biggl[\phi R-V(\phi)\biggr]+S_\mathrm{M}[g_{\mu\nu},\psi] \, ,
\label{eq:azione_Jordan}
\end{equation}
where we have defined the scalar field as $\phi \equiv f_R$ and the scalar field potential $V(\phi) \equiv \phi R(\phi)-f(R(\phi))$, if $f''(R) \neq 0$. The potential $V(\phi)$ is fixed by the specific functional expression of $f(R)$. Here, the scalar field is non minimally coupled to gravity trough the Ricci scalar.

Starting from the Jordan frame action in Eq.~\eqref{eq:azione_Jordan}, one can perform the conformal transformation of the metric considering \cite{Sotiriou:2008fR, Brax:2008hh}
\begin{equation}
    \tilde{g}_{\mu\nu}=\phi \, g_{\mu\nu} \;,
    \label{eq:trasf_conf_metric}
\end{equation}
where all the quantities with a tilde are those defined in the Einstein frame hereafter. Considering also a redefinition of the scalar field such that the new one $\tilde{\phi}$ couples minimally to the Ricci scalar and has canonical kinetic energy, i.e. 
\begin{equation}
\phi=e^{\sqrt{\frac{2}{3}} \kappa \tilde{\phi}} \; ,
\end{equation}
finally the action can be rewritten as
\begin{equation}
    \begin{aligned}
    S_E[\tilde{g}_{\mu\nu},\tilde{\phi}] = &\int d^4x\sqrt{-\tilde{g}}\left[\frac{\tilde{R}}{2\kappa^2}-\frac{1}{2}\tilde{g}^{\mu\nu}\partial_\mu\tilde{\phi}\partial_\nu\tilde{\phi}-V_\mathrm{E}(\tilde{\phi})\right] \\ &+S_\mathrm{M}[e^{-\sqrt{\frac{2}{3}} \kappa \tilde{\phi}}\tilde{g}_{\mu\nu},\psi] \, ,
    \end{aligned}\label{eq:Einstein_frame_action}
\end{equation}
where the Einstein frame potential is fixed by the original function $f(R)$ and is given by
\begin{equation}
    V_\mathrm{E}(\tilde{\phi})=\frac{Rf_R-f}{2\kappa^2 f_R^2} \, ,
\end{equation}
where $R=R(\tilde{\phi})$. We note that in this frame the matter source component is not minimally coupled with gravity anymore and the stress-energy tensor is also not conserved giving \cite{Faraoni:2004pi, Brax:2008hh}

\begin{equation}
    \tilde{\nabla}_{\mu}\tilde{T}^{\mu}_{\nu}=\frac{\kappa}{\sqrt{6}}\tilde{T}\,\tilde{\nabla}_{\nu}\tilde{\phi} \;,
    \label{eq:no_cons_Tmunu_Eframe}
\end{equation}
that means that in the Einstein frame matter feels a ``fifth" force acting on massive, non relativistic particles. Photons are an exception since the electromagnetic action is at the classical level conformally invariant or equivalently $\tilde{T}=0$ for radiation \citep{Brax:2008hh, Burrage:2017qrf}. 

\subsection{Theoretical viability constraints} \label{sec:theory_checks}
A generic $f(R)$ model does not automatically provide a consistent alternative to GR. In order to describe the late-time accelerated expansion while remaining compatible with local gravity tests and with the standard cosmological evolution at early times, the function $f(R)$ has to satisfy a number of theoretical and phenomenological viability conditions.

The first set of requirements concerns the absence of pathologies that would lead to changes in the nature of the gravitational interaction or instabilities of the non minimally coupled scalar field. In $f(R)$ gravity, looking at the Friedmann equation (Eq.~\eqref{eq:Friedmann_mod}), we see that the effective gravitational coupling is proportional to $G_\mathrm{N}/f_R$. Therefore, in order to preserve the attractive character of gravity and avoid a ghost-like graviton, one has to require \cite{Sotiriou:2008fR, DeFelice:2010aj, Appleby:2009uf}
\begin{equation}
    f_R>0 \,,
    \label{eq:f_R_positive}
\end{equation}
that means that the $f(R)$ function must be crescent with the $R$ and that the scalar field is an increasing function going back in the past. \\
The second condition is
\begin{equation}
    f_{RR}>0 \; ,
    \label{eq:f_RR_positive}
    \end{equation}
which ensures the stability of the scalar degree of freedom associated with $f(R)$ gravity and prevents the Dolgov-Kawasaki instability \cite{Dolgov:2003px}, that would correspond to a rapid growth of curvature perturbations in matter-dominated or high-density regions, making the theory incompatible with stable gravitational systems.

The same condition is closely related to the requirement that the scalaron is not tachyonic, i.e. has a negative squared mass. Around a constant-curvature background, such as a de Sitter or quasi-de Sitter solution, the effective scalaron mass is given by \cite{Sotiriou:2008fR}
\begin{equation}
    \mu^2_{\rm eff}=\frac{f_R-Rf_{RR}}{3f_{RR}} \; .
    \label{eq:squared_mass}
\end{equation}
Stability of the background requires that this term be positive in order to guarantee that small perturbations of the Ricci scalar do not grow exponentially around the background solution.

Further insight into the cosmological viability of a given model can be obtained by introducing the dimensionless quantities \cite{Amendola:2006we, Martinelli_2009}
\begin{equation}
\widetilde m(r)=\frac{Rf_{RR}}{f_R}, \qquad \text{where} \qquad r= -\frac{Rf_R}{f} \;.
\end{equation}
A model possesses a standard matter dominated era followed by a late-time accelerated phase if the function $m(r)$ satisfies
\begin{equation}
    \widetilde m(r) \simeq 0 \,, \qquad \widetilde m'(r)>-1 \qquad \text{for} \qquad r\simeq -1 \;.
    \label{eq:matter_condition}
\end{equation}

A further condition comes from local gravity constraints requiring a screening mechanism to not violate the Solar System and laboratory constraints \cite{Will:2014kxa, Dyson:1920cwa, Perielio-Mercurio}. Since the extra scalar degree of freedom written in the Einstein frame mediates an additional force, i.e. the ``fifth force" generated from the non conservation of the matter stress-energy tensor in Eq.~\eqref{eq:no_cons_Tmunu_Eframe}, this in principle would generally violate Solar System and laboratory constraints. Viable models must therefore implement a screening mechanism, the so-called chameleon mechanism \cite{Brax:2008hh}, where the scalaron mass depends on the ambient matter density: the field becomes heavy in high-density environments, suppressing deviations from GR, while it can remain light on cosmological scales.

A useful set of sufficient conditions is
\begin{equation}
V^{\prime}_\mathrm{E}(\tilde{\phi})=\frac{1}{\sqrt{6} \kappa f_R^2}\left(Rf_R-2f\right)<0 \; ,
\end{equation}
\begin{equation}
V^{\prime \prime}_\mathrm{E}(\tilde{\phi})=\frac{1}{3}\left[\frac{R}{f_R}+\frac{1}{f_{RR}}-\frac{4f}{f_R^2}\right]>0 \; ,
\end{equation}
and
\begin{equation}
V^{\prime \prime \prime}_\mathrm{E}(\tilde{\phi})=\frac{2 \kappa}{3\sqrt{6}}\left[\frac{3}{f_{RR}}+\frac{f_R f_{RRR}}{f_{RR}^3}+\frac{R}{f_R}-\frac{8f}{f_R^2}\right]<0 \; .
\end{equation}
These conditions express, respectively, the monotonic behavior of the potential, the positivity of the scalaron mass squared, and the behavior required for the chameleon mass to increase in dense environments \cite{Brax:2008hh}.

A viable late-time model must also recover GR in the regimes where it is well tested. In particular, in high-density environments and at curvatures much larger than the present cosmological curvature scale, the theory should approach GR with, at most, an effective cosmological constant. This requires
\begin{equation}
|f(R)-R| \ll R,
\qquad
|f_R-1| \ll 1,
\qquad
|Rf_{RR}|\ll 1 ,
\label{eq:newtonian_limit}
\end{equation}
in the appropriate high-curvature regime.

The above screening mechanisms allow us to not consider the Solar System and astrophysical bounds.

\subsection{Hu - Sawicki model}
Several functional forms of late-time $f(R)$ have been proposed in order to explain the cosmic expansion satisfying the above listed theoretical conditions. In this work, we will consider the Hu - Sawicki (HS) class of models \cite{Hu:2007nk} as a reference function to explore perturbations around with the GA. The HS correction to the EH term is written as 
\begin{equation}
    F(R)= - m^2\frac{c_1\left(R/m^2\right)^n}{c_2\left(R/m^2\right)^n+1} \;,
    \label{eq:fR_HS}
\end{equation}
where $m^2 = \Omega_\mathrm{m,0}H_0^2$, and $c_1$, $c_2$ and $n$ are model parameters. In particular, $n$ controls the rate at which the model approaches the $\Lambda\mathrm{CDM}$ limit in the high-curvature regime and from local gravity and equivalence-principle tests \cite{Capozziello:2007eu} results that $n > 0.9$. Moreover, following Ref.~\cite{fu_2010}, $n$ is often restricted to integer values.

The same expression can be recast and written as \cite{Bamba:2012qi, Basilakos:2013nfa} 

\begin{equation}
    F(R) =- \frac{2 \Lambda}{1+\left(\frac{b \Lambda}{R}\right)^n} \;,
    \label{eq:F_HS_b}
\end{equation}
where $\Lambda = (m^2 c_1)/(2c_2)$ and $b=2c_2^{1-1/n}/c_1$. Thus, the parameter $b$ determines how close the model is to $\Lambda\mathrm{CDM}$, in particular for $b \rightarrow 0$, then we have $F(R) \rightarrow-2\Lambda$. In general, the parameter $b$ can be related to $f_{R_0}$, which is more common in the literature, by taking the limit $f_{R_0} = 1 + F_{R|_{R\rightarrow R_0}}$ in the previous equation and working to linear order at late times $(a=1)$.

It is easy to check that $\lim_{R\to\infty} F(R) = -2 \Lambda $, so that the HS model can reproduce a $\Lambda$CDM-like expansion. Moreover, as shown in \citet{Hu:2007nk}, the model satisfies the theoretical viability requirements discussed above.

\section{Genetic Algorithms}\label{sec:Genetic_algorithms}
Genetic Algorithms (GAs) are a class of machine learning (ML) techniques that implement a stochastic search in a space of possible solutions. They are used as a symbolic regression method, i.e. as a procedure to reconstruct analytic functions directly from data, without imposing a specific parametric ansatz in advance. 
Here, we provide an overview of the general mechanism, while we refer to \cite{Bogdanos:2009ib, Nesseris:2010ep} for a more detailed description and to \cite{Nesseris:2012tt, Nesseris:2013bia, Arjona:2019fwb, Arjona:2020kco, Arjona:2021hmg, Aizpuru:2021vhd, Orjuela-Quintana:2022nnq, Kamerkar:2022dfu, Gangopadhyay:2023nli, Orjuela-Quintana:2023uqb, Orjuela-Quintana:2024hha, Arjona:2024dsr} for cosmological application of GAs. 

The basic idea of a GA is inspired by biological evolution and natural selection. The algorithm starts from an initial population of candidate functions ($N_{\rm pop}$), which is randomly generated providing only a seed. Candidate functions are defined according to a predefined grammar that specifies the elementary building blocks from which the functions can be constructed, such as constants, powers, polynomials, etc. The choice of the grammar and of the population size has already been extensively tested in previous analyses \cite{Bogdanos:2009ib}. Each individual in the population is then evaluated according to a fitness criterion, which measures how well the function describes the data and is quantified through a $\chi^2$ statistic. The selection probability is then chosen so as to favor the individuals with better fitness, without making the process fully deterministic. 
Once the fitness has been evaluated, the functions with the minimum $\chi^2$ are selected and used to generate a new population through the genetic operations of crossover and mutation. 

Crossover combines parts of two or more parent functions to produce new offspring, while mutation randomly modifies part of an existing function. These operations are crucial in the GA algorithm since they introduce variation in the population and allow the algorithm to explore new regions of the functional space. At the end of each generation, the newly generated functions replace partially the previous population, maintaining the total initial number $N_{\rm pop}$ of individuals.
The procedure is then iterated until a convergence criterion is satisfied or a fixed maximum number of generations is reached. In this way, the population evolves generation after generation, progressively favoring analytic functions that provide a better fit to the data.

In this work, we apply this procedure to reconstruct the functional form of the $f(R)$ function that best fits the background late-time cosmological data. Since our aim is to explore smooth deviations around the HS model, and since the numerical solution of the $f(R)$ background equations is computationally demanding, we adopt a restricted grammar made only of polynomial functions. This choice reduces the complexity of the reconstructed functions and helps avoid numerical instabilities associated with more general symbolic expressions.
The initial population is composed of $N_{\mathrm{pop}}=100$ individuals, while the number of generations is set to $N_{\mathrm{gen}}=10$. 
The initial functions are generated from a fixed seed, which determines the starting population of the GA. 

In principle, several independent seeds should be tested to assess whether the reconstruction depends on the initial population. We therefore performed a preliminary analysis using different seeds and compared the final $\chi^2$ values obtained from populations generated from different initial conditions. We found that, independently of the initial population and hence of the seed, the final $\chi^2$ remains stable at the last iteration. This also suggests that the adopted number of individuals in the initial population and the number of generations are sufficient for the purposes of our analysis. 

Also, the choice of $N_{\rm gen}$ and $N_{\rm pop}$ results from a compromise between allowing the population to evolve sufficiently and limiting the high computational cost of each fitness evaluation, which requires solving the background system in Eq.~\eqref{eq:y_ODE}. The final output of the algorithm is therefore a smooth analytic function $f_{\rm GA}(R)$, which provides the GA reconstruction of the underlying $f(R)$ model.

\section{Dataset and Methodology}\label{sec:dataset_methodology}
The goal of this work is to constrain smooth deviations around the Hu--Sawicki (HS) model using current background cosmological data. In this section, we describe the observational datasets used in the analysis and the methodology adopted to compare the reconstructed $f(R)$ functions with data.

\subsection{Cosmological data} \label{sec:data}
We aim to test whether current background observations are compatible with a cosmological evolution close to the HS scenario. To this end, we use two complementary probes of the late-time expansion history: Baryon Acoustic Oscillations (BAO) and Type Ia supernovae (SNe). 

\subsubsection{Baryon Acoustic Oscillations}BAO provide a standard ruler in cosmology, corresponding to the comoving sound horizon at the baryon drag epoch, $r_d$. This characteristic scale is imprinted in the large-scale clustering of matter and allows one to constrain cosmological distances relative to $r_d$. In the transverse direction, BAO measurements probe the comoving angular diameter distance $D_M(z)=(1+z)d_A(z)$, while in the radial direction they constrain the Hubble distance $D_H(z)\equiv c/H(z)$. BAO analyses usually express these constraints through scaling parameters, which quantify the geometrical distortion of the measured clustering signal with respect to a fiducial cosmology. In particular, the transverse and radial Alcock--Paczynski parameters are defined as
\begin{equation}
\alpha_\perp =
\frac{D_M(z)/r_d}{D_M^{\rm fid}(z)/r_d^{\rm fid}},
\qquad
\alpha_\parallel =
\frac{D_H(z)/r_d}{D_H^{\rm fid}(z)/r_d^{\rm fid}}.
\end{equation}
From these quantities one can define the isotropic dilation parameter
\begin{equation}
\alpha_{\rm iso} \equiv 
\left(\alpha_\perp^2 \alpha_\parallel \right)^{1/3}
= \frac{D_V(z)/r_d}{D_V^{\rm fid}(z)/r_d^{\rm fid}},
\end{equation}
where
\begin{equation}
D_V(z)=\left[zD_M^2(z)D_H(z)\right]^{1/3},
\end{equation}
and the anisotropic Alcock--Paczynski parameter
\begin{equation}
\alpha_{\rm AP} \equiv 
\frac{\alpha_\parallel}{\alpha_\perp}
= \frac{D_H(z) D_M^{\rm fid}(z)}{D_H^{\rm fid}(z) D_M(z)}.
\end{equation}
While $\alpha_{\rm iso}$ measures the overall isotropic rescaling of the BAO feature, $\alpha_{\rm AP}$ captures the relative distortion between the radial and transverse directions. In these expressions, the superscript \textit{fid} denotes the fiducial cosmology adopted. Since the BAO feature is measured after mapping observed angles and redshifts into distances using an assumed cosmology, the parameters $\alpha_\perp$ and $\alpha_\parallel$ quantify the rescaling needed to transform the fiducial transverse and radial distances into those of the cosmological model being tested. In our work we adopted the same fiducial cosmology as DESI DR2 \citep{DESI:2025zgx}.

In this work, we use the BAO measurements from DESI DR2 \cite{DESI:2025zgx}, implemented in our likelihood through the compressed BAO variables $\alpha_{\rm iso}$ and $\alpha_{\rm AP}$. The DESI DR2 sample spans the effective redshift range $0.295 \leq z \leq 2.330$ and includes measurements from different large-scale structure tracers, namely BGS, LRGs, ELGs, QSOs, and the Ly$\alpha$ forest. Since BAO measurements constrain distances relative to the sound horizon scale, we fix the physical baryon density to the BBN value of \cite{Schoneberg_2024} $\Omega_bh^2=0.02218$ in order to calibrate $r_d$ and extract constraints on the late-time expansion history. 

This choice is made for computational reasons, since allowing $\Omega_bh^2$ to vary with a BBN prior would significantly increase the computational cost. However, we acknowledge that this is a simplifying assumption, as a more complete treatment would require sampling $\Omega_bh^2$ with a BBN prior, thereby introducing an additional source of uncertainty in the inferred late-time cosmological constraints.

\subsubsection{Supernovae}

We also include SNe, which provide measurements of the distance modulus $\mu(z)$, defined as the difference between the apparent magnitude $m_B(z)$ and the absolute magnitude $M_B$. In particular, we use the \textit{Pantheon+} catalogue, which contains 1701 SNe Ia in the redshift range $0.001 \leq z \leq 2.26$, drawn from 18 different surveys \citep{Scolnic:2021amr}. The apparent magnitude is related to the luminosity distance through
\begin{equation}
m_B(z) = M_B + 5 \log_{10} \left(\frac{d_L(z)}{\text{Mpc}}\right) + 25,
\label{eq:mz}
\end{equation}
where $d_L(z)$ encodes the cosmological information, while $M_B$ acts as a calibration parameter.

A key point is that $M_B$ is degenerate with the Hubble constant $H_0$, unless external calibration information is included. The Pantheon+ sample can be calibrated using the SH0ES distance ladder \citep{SH0ES}. In this approach, geometrical distance indicators are used to calibrate Cepheid variables in the Milky Way and in nearby galaxies. These Cepheids then calibrate the absolute magnitude of SNe Ia in the same host galaxies, providing a direct determination of $M_B$ and therefore fixing the absolute distance scale.

In this work, however, we do not impose an external calibration on $M_B$. Therefore, the SNe constrain only relative distances and do not provide an independent absolute calibration of $H_0$. We account for this by marginalizing over the supernova nuisance calibration, equivalently over the magnitude offset associated with the $M_B$--$H_0$ degeneracy, following the procedure described in Appendix C of \cite{Conley_2011}.

\subsection{Overview of the complete algorithm}\label{sec:methodology}
\begin{figure}
    \centering
    \includegraphics[width=0.95\linewidth]{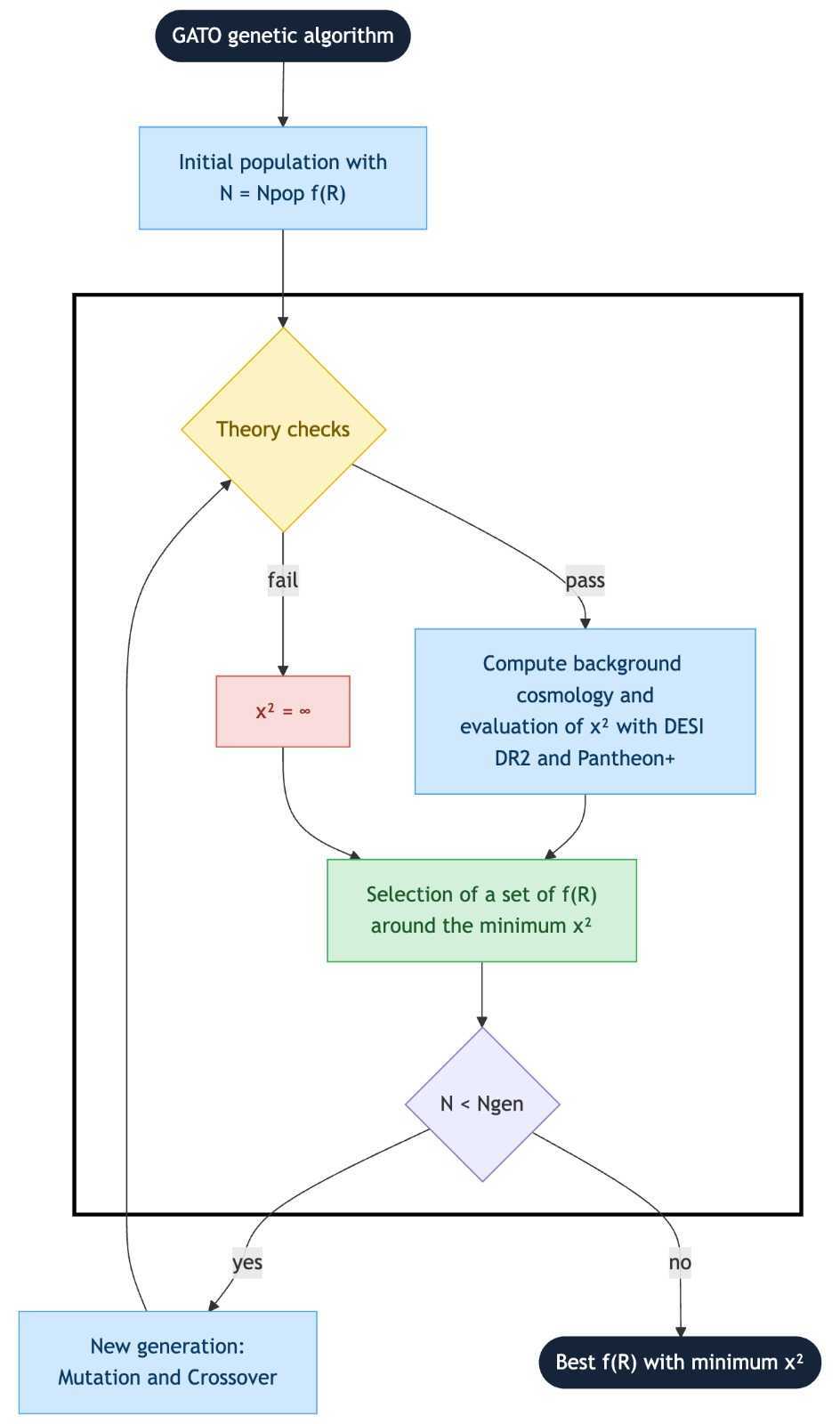}
    \caption{Flow chart of the GA and sampling process.}
    \label{fig:flow_chart}
\end{figure}

To test small deviations around the HS model, we reconstruct the functional form of $F(R)$ starting from the expression given in Eq.~\eqref{eq:F_HS_b}, and employing the Genetic Algorithms (GA) introduced in Section~\ref{sec:Genetic_algorithms} and based on the \texttt{GATO}\footnote{\url{https://github.com/snesseris/Genetic-Algorithms}}  code \cite{Bogdanos:2009ib, Nesseris:2010ep, Nesseris:2012tt, Nesseris:2013bia, Sapone:2014nna, Arjona:2019fwb, Arjona:2020kco, Hogg:2020ktc, Martinelli:2020hud, Renzi:2020bvl, Arjona:2020axn, Kamerkar:2022dfu}.
Once the GA provides a candidate $F(R)$, this function is passed to our pipeline, based on the \texttt{CANDI}\footnote{\url{https://github.com/chiaradeleo1/CANDI}} cosmology code, which computes the corresponding cosmological evolution and compares it with the data. In the following, we describe the steps implemented in the pipeline, while a schematic representation of the workflow is shown in Fig.~\ref{fig:flow_chart}. As discussed above, this restriction is motivated by the high computational cost of solving the full $f(R)$ background equations and by the fact that fully free GA-generated functions are in general not guaranteed to lead to viable and numerically stable cosmological evolutions.

As discussed in Section~\ref{sec:Genetic_algorithms}, the GA starts from the generation of an initial population of $N_{\rm pop}=100$ individuals, which is then evolved for $N_{\rm gen}=10$ generations. This set of individuals is generated around a prior \textsc{GA\_prior} based on Eq.~\eqref{eq:F_HS_b}, which is defined as
\begin{equation} 
\text{\textsc{GA\_prior}}=-\frac{2\Lambda}{1+\mathrm{GA_x}} \,, \label{eq:GA_prior} 
\end{equation}
where $\mathrm{GA_x}$ denotes the symbolic degree of freedom evolved by the GA. This quantity is built as a combination of elementary functions selected from the GA grammar and can be written in the generic form

\begin{equation}
    \mathrm{GA_x}=\sum_i \gamma_{1,i}\left[g_i(\gamma_{3,i}x)\right]^{\gamma_{4,i}} \,,
    \label{eq:GAx_ci}
\end{equation}
where $x$ denotes the input variable of the GA, that in our case is the inverse of the Ricci scalar $\Lambda/R$. The function $g_i$ is selected from the predefined grammar through the discrete coefficient $\gamma_{2,i}$. More explicitly, if we define a list of possible grammar functions, for example \texttt{[const, poly, cpl]}, then $\gamma_{2,i}$ selects one element of this list. In our case, $\gamma_{2,i}$ is fixed because the only allowed grammar is the polynomial one. This restriction is motivated by the aim of this work, which is to study small deviations around the HS model, without exploring completely arbitrary functional forms. The coefficients $\gamma_{1,i}$ and $\gamma_{3,i}$ are randomly picked in the range $[-0.5,0.5]$, while $\gamma_{4,i}$ is an integer taken in the range $[0,5]$ and corresponds to the polynomial mode $n$.

Each $F(R)$ generated in the initial population must pass the theoretical viability checks described in Section~\ref{sec:theory_checks} before entering the full pipeline. If any of these checks fails, the corresponding $F(R)$ is discarded by assigning $\chi^2=\infty$; while if the candidate passes all theory checks, it is used as input to compute the corresponding cosmology.
\noindent
This computation is performed through the \texttt{CANDI} theory module \cite{DeLeo:2025rhy, Fazzari:2025nfr}. In particular, we implement an additional expansion model that computes $H(z)$ in the $f(R)$ scenario by solving the system of differential equations described in Eq.~\eqref{eq:y_ODE}. The output of the theory module is therefore the Hubble expansion history associated with the reconstructed $F(R)$.

Once the background cosmology has been obtained, it is used as the theoretical prediction in the likelihood evaluation. The comparison with the data is performed within the \texttt{CANDI} likelihood framework. As described in Section~\ref{sec:data}, we compare the reconstructed background cosmology against BAO and SNe data. Since solving the $f(R)$ background equations is computationally expensive, we do not use an external sampler for each GA individual. Instead, we adopt a grid-based minimization procedure, which is sufficient for assigning a fitness value, namely the minimum $\chi^2$, to each candidate $F(R)$.
\noindent
In practice, for each viable $F(R)$ we construct a grid in $H_0$ and $\Omega_m$, while fixing $\Omega_b h^2$ to the BBN value and marginalizing over $M_B$ following the procedure described in Appendix C of \cite{Conley_2011}. Hereafter we adopt the notation $\Omega_{m}\equiv \Omega_{m,0}$. 

The grid search is performed within the ranges $50 \leq H_0 \leq 80$ and $0.2 \leq \Omega_m \leq 0.45$. For each value of $H_0$, we scan over $\Omega_m$ and select the value that minimizes 
\begin{equation}
\left|H(z_{\min}) - H_0\right|,
\end{equation}
where $H(z)$ is the reconstructed Hubble expansion obtained from the $f(R)$ background equations, i.e. Eqs. \eqref{eq:y_ODE}. 
The minimization is carried out in three successive steps. We first perform a broader search using a grid of 8 points in $H_0$ and 24 points in $\Omega_m$. We then construct a second, refined grid centered around the best point found in the first step, and finally perform a third zoomed search around the minimum obtained in the second step. The likelihood is evaluated on the final resulting pair, and the corresponding $\chi^2$ is computed for the best-fit values of $H_0$ and $\Omega_m$ associated with the reconstructed $F(R)$. This value of $\chi^2$ is then returned to the GA and used as the fitness of that candidate function.

This procedure is repeated for all the functions in the initial population, so that each candidate $F(R)$ is assigned a corresponding $\chi^2$. The new generation is then constructed by selecting the best-performing fraction of the population, defined in the code by the parameter \texttt{selectionrate}. In our case, this corresponds to the $30\%$ of functions with the lowest $\chi^2$ values. Starting from these selected candidates, the next generation is built through the mutation and crossover operations described in Section~\ref{sec:Genetic_algorithms}.

The newly generated population then undergoes the same sequence of steps: theoretical viability checks, computation of the background cosmology, likelihood evaluation, and assignment of a fitness value through the corresponding $\chi^2$. This iterative procedure is repeated for all $N_{\rm gen}=10$ generations. From the final generation, we identify the overall best-fit $F(R)$, together with its associated minimum $\chi^2$ and best-fit cosmological parameters. Since the GA does not directly provide an uncertainty on the reconstructed function, we estimate an empirical error region from the distribution of viable $F(R)$ functions in the final generation. 

In particular, we select the subset of functions corresponding to the lowest $68\%$ of the $\chi^2$ distribution around the minimum, in analogy with the $68\%$ confidence region of a Gaussian likelihood. The envelope spanned by these functions is then used to define the uncertainty band associated with the best-fit $F(R)$, an approach similar to what was done in \cite{Kamerkar:2022dfu}. Also, we should note that, while in general the 68\% envelope is a good proxy for the confidence regions, in some cases if the population  of functions is not very dense, this could introduce some bias, so some care might be needed when itnerpreting the results.

\section{Results}\label{sec:results}
The aim of this work is to constrain smooth deviations around the Hu--Sawicki (HS) model using current background cosmological data, namely BAO measurements from DESI DR2 \citep{DESI:2025zgx} and SNe from the Pantheon+ catalogue \citep{Scolnic:2021amr} as discussed in Section~\ref{sec:data}. These deviations are generated with the \texttt{GATO} Genetic Algorithm tool, as described in Section~\ref{sec:Genetic_algorithms}, from which we obtain a final best-fit $F(R)$.

We present here the results obtained following the methodology discussed in Section~\ref{sec:methodology}. Specifically, from the full pipeline we find that the best-fit $F(R)$, written in the form described in Eq.~\eqref{eq:F_HS_b}, is
\begin{equation}
F(R)=
-\dfrac{2\Lambda}
{1+\left(\dfrac{1.12 \times 10^{-3}\,\Lambda}{R}\right)^2
+\left(\dfrac{1.44 \times 10^{-2}\,\Lambda}{R}\right)^3} \, ,
\label{eq:best_fit_FR}
\end{equation}
where the best-fit cosmological parameters are $H_0=68.698\;{\rm km\,s^{-1}\,Mpc^{-1}}$, $\Omega_m=0.306$, and $\Lambda = 3H_0^2(1-\Omega_m)$. From this expression, we note that the correction is dominated by the $n=2$ term.
\begin{figure*}[t]
    \centering
    \includegraphics[width=0.8\linewidth]{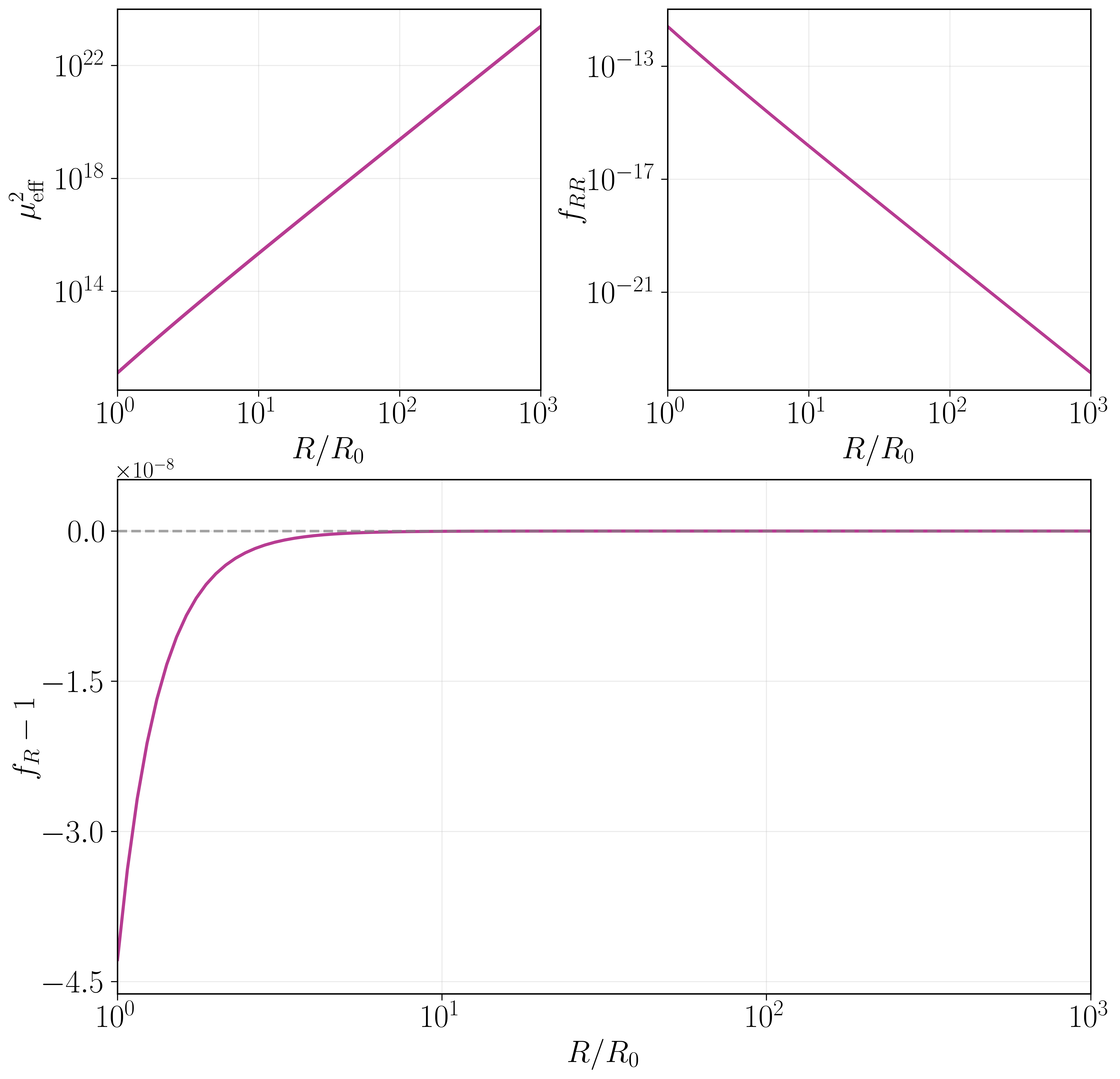}
    \caption{Theoretical viability checks for the best-fit $F(R)$ from GA reconstruction as a function of the normalized curvature $R/R_0$. Top left: squared effective mass $\mu^2_{\rm eff}$. Top right: second derivative $f_{RR}$. Bottom: deviation of the first derivative from the GR limit, $f_R-1$. The dashed gray line marks the GR limit $f_R-1=0$.}
    \label{fig:theory_checks}
\end{figure*}

Before discussing the behaviour and the physical implications of the best-fit reconstructed $F(R)$, we first verify that it satisfies the theoretical viability conditions. In Fig.~\ref{fig:theory_checks}, we show the effective lasquared mass $\mu^2_{\rm eff}$, defined in Eq.~\eqref{eq:squared_mass}, the second derivative $f_{RR}$, and the quantity $f_R-1$, where we recall that $f(R)=R+F(R)$ and $f_R$ is the first derivative defined in Eq.~\eqref{eq:fR_FR}. The reconstructed model satisfies the expected stability requirements over the full curvature range considered: both $\mu^2_{\rm eff}$ and $f_{RR}$ remain positive, avoiding tachyonic and Dolgov--Kawasaki instabilities, respectively. 

Moreover, the deviation of $f_R$ from its GR value is extremely small, with $|f_R-1|\ll 1$ throughout the entire interval. In particular, the lower panel shows that at lower values of $R$ the deviation remains of order $10^{-8}$ while rapidly approaches zero as the curvature increases. This behaviour is consistent with previous results, such as \cite{Martinelli_2009, Cataneo:2014kaa, Liu:2016xes}, where viable HS-like models are found to require deviations of order $10^{-7}$-$10^{-8}$ around today. The simultaneous increase of $\mu^2_{\rm eff}$ and decrease of $f_{RR}$ at large curvature further indicates that the extra scalar degree of freedom becomes increasingly heavy, so that modifications to GR are efficiently suppressed in the high-curvature regime. 

\begin{figure*}[t]
    \centering
    \includegraphics[width=0.8\linewidth]{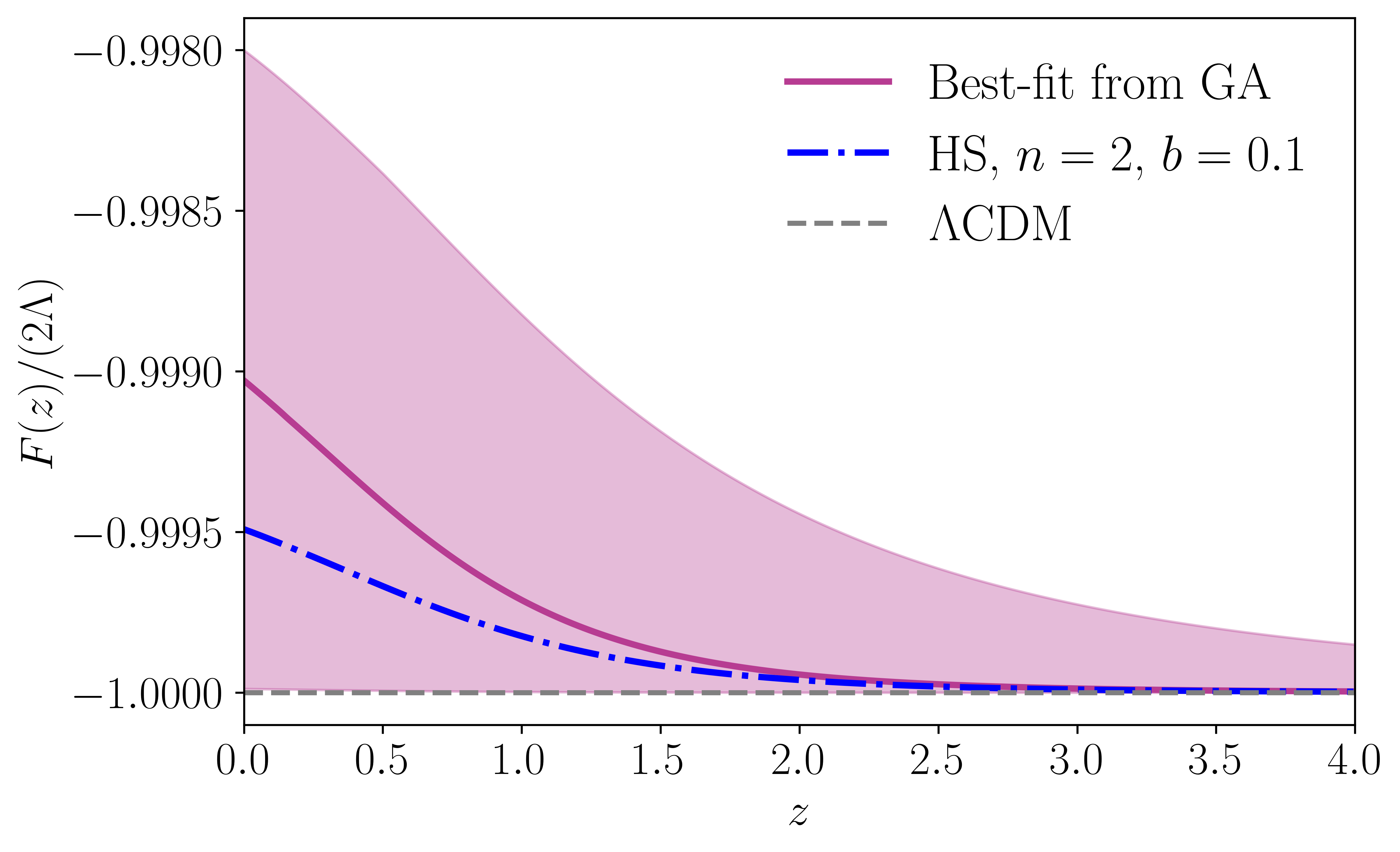}
    \caption{Best-fit reconstructed $F(R)$ function normalized by $2\Lambda$ as a function of redshift $z$. The solid purple line shows the best-fit GA reconstruction, while the dashed gray line denotes the $\Lambda$CDM limit, corresponding to $F(R)/(2\Lambda)=-1$, and the blue line shows the Hu--Sawicki limit for $n=2$ and $b=0.1$. The shaded purple region represents the empirical $68\%$ uncertainty band estimated from the final GA generation. The region below $F(R)/(2\Lambda)=-1$ is excluded by the theoretical viability checks discussed in the text.}

    \label{fig:fR}
\end{figure*}

In Fig.~\ref{fig:fR}, we show the best-fit reconstructed $F(R)$ obtained from the GA pipeline, normalized by $2\Lambda$, as a function of $z$, where the redshift dependence is encoded through the background Ricci scalar $R(z)=6\left[2H^2(z)-(1+z)H(z)\,H'(z)\right]$. The dashed gray line represents the $\Lambda$CDM limit, corresponding to $F(R)/(2\Lambda)=-1$, while the blue line shows the HS model with $n=2$ and $b=0.1$, used as a reference model following the formulation of Eq.~\eqref{eq:F_HS_b}. Finally, the shaded purple region shows the empirical $68\%$ uncertainty band estimated from the final GA generation, as described above. 

The value $n=2$ is commonly adopted in the literature and is particularly useful here because the GA reconstruction selects a deformation dominated by a quadratic correction in $\Lambda/R$. On the other hand, the parameter  $b=0.1$ is chosen even though is much larger than the currently allowed value of $f_{R_0}\sim 10^{-6}$ when all current data, including cosmic microwave background (CMB), BAO, SNe and others are taken into account, as a proxy/upper limit of of where the deviations from $\Lambda$CDM become extremely strong, as the parameter is close to $O(1)$ \cite{Basilakos:2013nfa}.

The reconstructed GA function displays the same qualitative behavior as the HS reference model. In particular, it approaches the $\Lambda$CDM value from above and does not cross below $F(R)/(2\Lambda)=-1$. Having defined the coefficients $\gamma_{1,i}$ and $\gamma_{3,i}$ to vary in the interval $[-0.5,0.5]$, as given in Eq.~\eqref{eq:GAx_ci}, the GA can in principle generate candidates with $F(R)/(2\Lambda)<-1$. However when the negative contributions in the denominator become dominant, the resulting functions typically violate the stability requirements, in particular $f_{RR}>0$ and $\mu^2_{\rm eff}>0$, and so they are discarded by the theory viability checks.

Compared with the HS curve with $n=2$ and $b=0.1$, the GA best-fit predicts a slightly larger deviation from $\Lambda$CDM at low redshift. This means that, within the freedom allowed by the data and by the viability conditions imposed in the pipeline, the GA favours a somewhat stronger late-time correction than the chosen HS reference model. However, this enhancement is still very small, of order $10^{-3}$, and remains inside the same smooth deformation regime. As redshift increases, the GA curve rapidly converges towards both the HS reference curve and the $\Lambda$CDM limit. This indicates that the reconstructed modification is efficiently suppressed at higher curvature, as expected for viable $F(R)$ models.

Physically, this behaviour means that the reconstructed deviation is concentrated in the low-curvature, late-time regime, where modified-gravity effects can contribute to the accelerated expansion. At higher redshift, corresponding to larger values of $R(z)$, the function becomes almost indistinguishable from the constant contribution $F(R)=-2\Lambda$. This agreement with the $\Lambda$CDM limit at high redshift is also enforced by construction: in the GA pipeline we impose the matching condition in the matter-dominated regime, so that the reconstructed model is required to recover the $\Lambda$CDM background when matter dominates. In this sense, the GA reconstruction inherits the main physical feature of the HS model: it allows a small late-time departure from $\Lambda$CDM while recovering the GR-like high-curvature regime.

We report in Table~\ref{tab:chi2} the values of the minimum $\chi^2$ obtained using DESI DR2 and Pantheon+, together with the corresponding best-fit values of $H_0$ and $\Omega_m$ obtained with the GA pipeline for the $\Lambda$CDM model, the HS and the reconstructed $F(R)$ model. In principle, it would also be interesting to compare these results with a CPL parametrization. However, our pipeline reconstructs the cosmic expansion starting from a given $F(R)$ function, and an equivalent $F(R)$ expression corresponding to the CPL model is not available in the literature. For this reason, we do not report CPL results here, since a consistent comparison should be performed within the same pipeline and under the same numerical assumptions.
In this comparison, $\Omega_b h^2$ is fixed to the BBN value, while $M_B$ is treated as a nuisance parameter and marginalized over. The three models give very similar values of $\chi^2$. In particular, the reconstructed $F(R)$ gives a slightly smaller value; however, these differences are small and do not indicate a statistically significant preference for one model over the others. By construction, the reconstructed model approaches the $\Lambda$CDM limit at high redshift, so the deviations with respect to the reference model are mainly confined to the low-redshift regime.

This result is consistent with the behavior shown in Figs.~\ref{fig:theory_checks} and~\ref{fig:fR}. The best-fit GA function remains very close to the $\Lambda$CDM limit and follows the same qualitative behaviour as the HS reference model. Therefore, at the level of background observables probed by DESI DR2 and Pantheon+, the reconstructed $F(R)$ model acts only as a mild deformation of $\Lambda$CDM rather than as a strongly distinct cosmological scenario. The small improvement in $\chi^2$ suggests that the additional functional freedom of the GA can slightly better accommodate the data, but the reconstructed function is still tightly constrained to remain close to the standard constant contribution $F(R)=-2\Lambda$.

We do not perform a formal model-selection analysis based on information criteria such as AIC, BIC or DIC \citep{wagenmakers2007}. The reason is that, in the GA framework, there is no unique and unambiguous definition of the number of model parameters: the reconstructed function is generated through symbolic expressions with variable structure, complexity, and effective degrees of freedom. Therefore, assigning a fixed parameter count to the GA reconstruction would be arbitrary. For this reason, the $\chi^2$ values in Table~\ref{tab:chi2} should be interpreted only as a goodness-of-fit comparison, rather than as a statistically rigorous model-selection criterion.
\begin{table}[h!]
    \centering
    \begin{tabular}{lccc}
        \hline
        Model & $\chi^2_{\rm min}$ & $H_0\;{\rm [km\,s^{-1}\,Mpc^{-1}}]$ & $\Omega_m$ \\
        \hline
        $\Lambda$CDM & $1419.5$ & $68.89$ & $0.3046$ \\
        HS, $n=2$, $b=0.1$ & $1420.1$ & $68.89$ & $0.2967$ \\
       $f_{\rm GA}(R)$ & $1418.9$ & $68.89$ & $0.3051$ \\
        \hline
    \end{tabular}
    \caption{Minimum $\chi^2$ values and corresponding best-fit parameters for $\Lambda$CDM, the Hu-Sawicki (HS) reference model with $n=2$ and $b=0.1$, and the best-fit GA reconstructed $F(R)$ model.}
    \label{tab:chi2}
\end{table}

\begin{figure*}[t!]
    \centering
    \begin{subfigure}{0.49\linewidth}
        \centering
        \includegraphics[width=\linewidth]{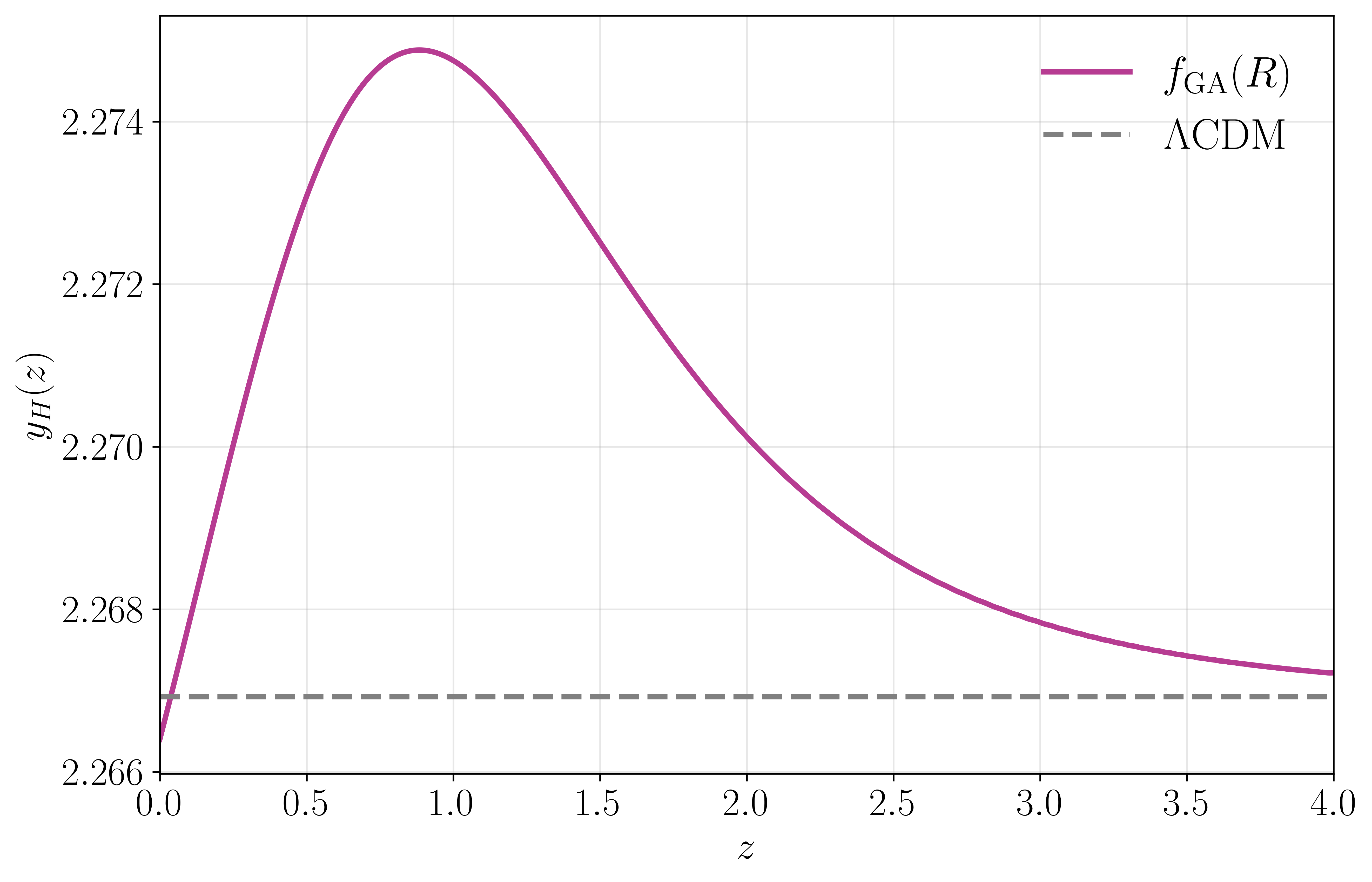}
        \caption{}
        \label{fig:yH}
    \end{subfigure}
    \hfill
    \begin{subfigure}{0.47\linewidth}
        \centering
        \includegraphics[width=\linewidth]{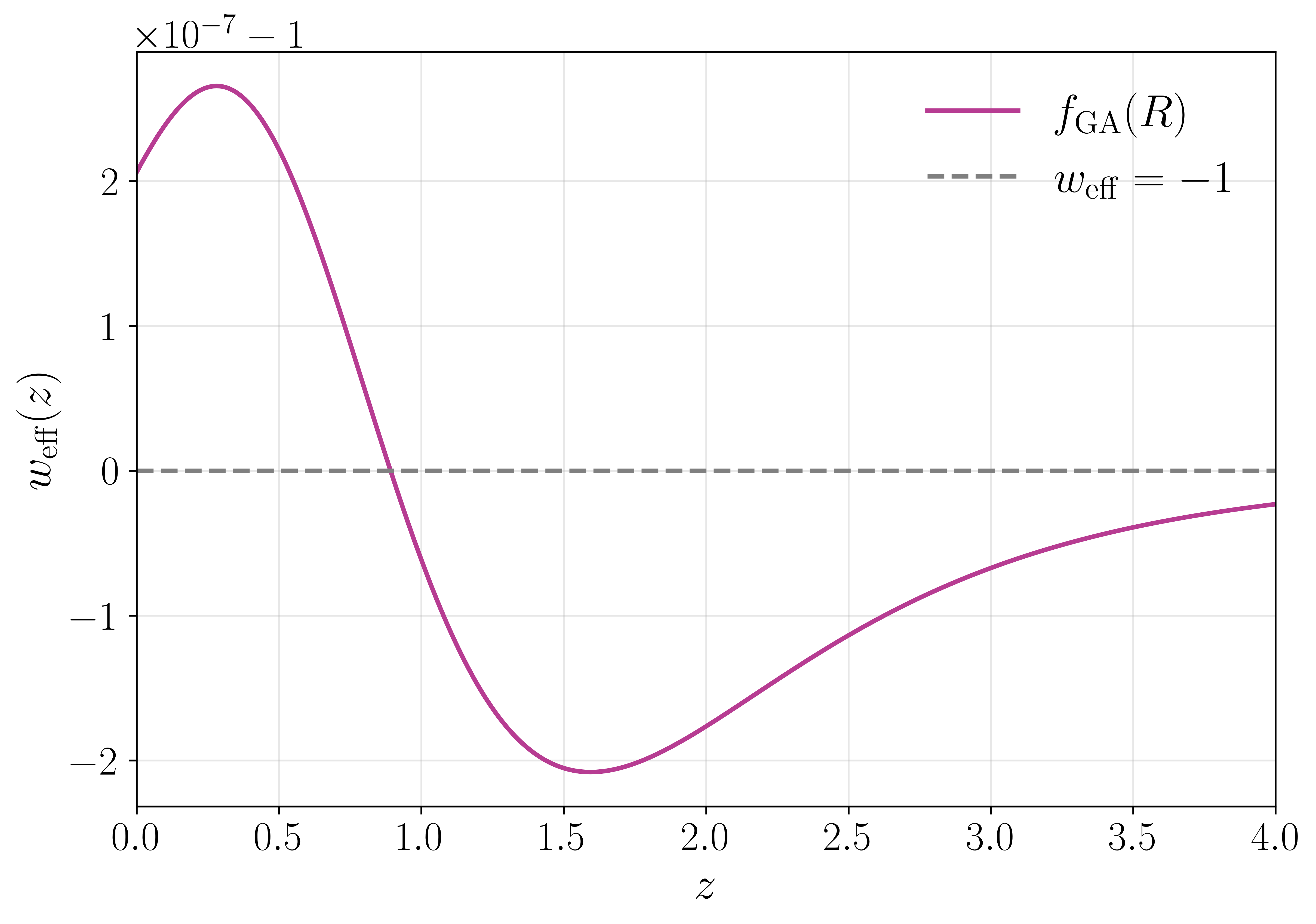}
        \caption{}
        \label{fig:weff}
    \end{subfigure}

    \caption{Evolution of the dimensionless variable $y_H(z)$ (left) and $w_{\rm eff}(z)$ (right) for the best-fit GA model $f_{\rm GA}(R)$ compared with the $\Lambda$CDM prediction in dashed gray. }
    \label{fig:yH_weff}
\end{figure*}

In Fig.~\ref{fig:yH}, we show the evolution of $y_H(z)$ for the best-fit GA model, compared with the corresponding $\Lambda$CDM prediction in gray. In $\Lambda$CDM, $y_H$ is constant, since the DE contribution is given by a pure cosmological constant. In contrast, the reconstructed $F_{\rm GA}(R)$ induces a mild redshift dependence in $y_H$, reflecting the curvature dependence of the effective DE sector.
At low redshift, the GA solution starts slightly below the $\Lambda$CDM value and crosses it around $z\simeq 0.1$. It then increases up to a maximum around $z\simeq 1$, before decreasing and progressively approaching the $\Lambda$CDM behaviour at higher redshift as required by construction.

We find these two characteristic redshift scales also in Fig.~\ref{fig:weff}, where we show the evolution of the effective EoS parameter $w_{\rm eff}(z)$. In particular, the low-redshift feature of $y_H$ is reflected in the small positive peak of $w_{\rm eff}$, while the maximum of $y_H$ around $z\simeq 1$ corresponds to the crossing of $w_{\rm eff}=-1$. This connection arises from the fact that $w_{\rm eff}$ is controlled by the redshift evolution of $y_H$, rather than by its absolute value. Indeed, the Eq.\eqref{eq:wDE_yH} can be rewritten in terms of redshift as
\begin{equation}
w_{\rm eff}(z)
=
-1+\frac{1+z}{3y_H}\frac{dy_H}{dz}.
\end{equation}
Therefore, when $y_H$ increases with redshift, one has $dy_H/dz>0$ and the effective EoS lies slightly above $-1$. 

Conversely, after $y_H$ reaches its maximum, $dy_H/dz$ changes sign and $w_{\rm eff}$ moves slightly below $-1$. The crossing of $w_{\rm eff}=-1$ therefore marks the turning point of $y_H$, while the amplitude of $w_{\rm eff}$ measures how rapidly $y_H$ varies with redshift. Physically, this shows that the reconstructed $F(R)$ induces only a very small dynamical evolution of the effective DE sector, with a mild transition from an effective quintessence-like behaviour to an effective phantom-like behaviour. 

Nevertheless, the deviation from $w_{\rm eff}=-1$ is extremely small, of order $10^{-7}$, indicating that this quintessence-to-phantom transition is only a tiny dynamical feature of the reconstruction and that the background expansion remains practically indistinguishable from $\Lambda$CDM. This behaviour is consistent with the trend shown in Fig.~3 of Ref.~\cite{Hu:2007nk}, where, for a fixed value of $n$, smaller absolute values of $F_{R_0}$ make the effective equation of state progressively closer to the $\Lambda$CDM limit. In our case, the small amplitude of the deviation from $w_{\rm eff}=-1$ therefore reflects the fact that the best-fit reconstruction lies very close to the $\Lambda$CDM regime. 

While it might be tempting to compare $w_{\rm eff}$ with the DESI results of \cite{DESI:2024mwx} on the CPL parameterization, in fact we cannot compare the different redshift behaviors of the two directly, given one is a Taylor expansion at late times \cite{Nesseris:2025lke} and the other a physical model covering the whole redshift range. Moreover, as shown in \cite{Hu:2007nk} the behavior of $w_{\rm eff}$ is vastly different compared to that of the CPL parameterization, as in the Hu-Sawicki model $w_{\rm eff}$ crosses -1 several times and asymptotes to -1 at high redshifts. 

\begin{figure*}[t]
    \centering
    \includegraphics[width=0.8\linewidth]{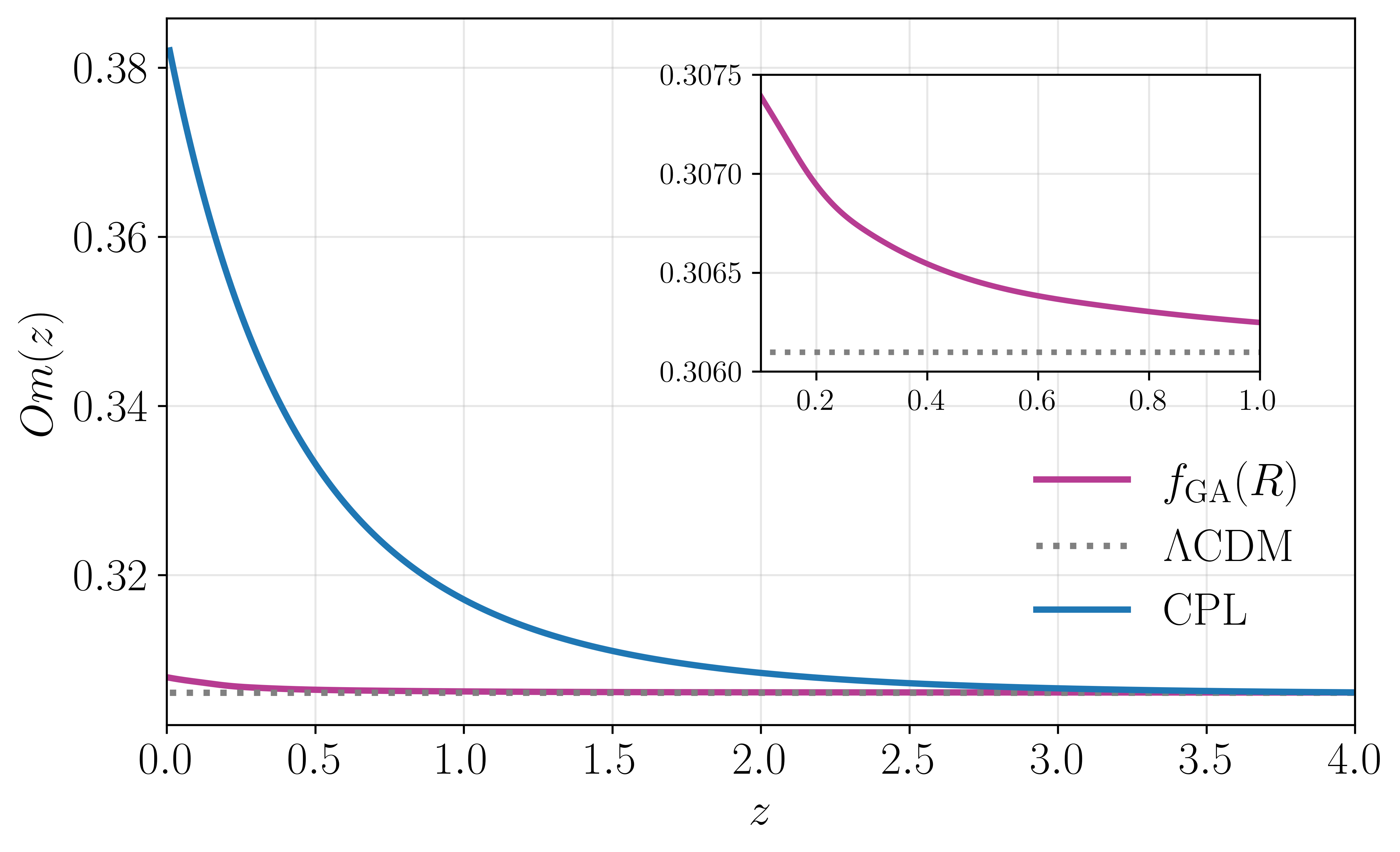}
    \caption{Evolution of the $Om(z)$ diagnostic for the best-fit GA reconstructed $F(R)$ model $f_{\rm GA}(R)$ in purple, compared with the $\Lambda$CDM prediction in dashed gray and the CPL parametrization fitted to the same dataset in blue.}
    \label{fig:Om_statistic}
\end{figure*}

Finally, we consider the $Om(z)$ diagnostic \citep{Sahni:2008xx}, defined as
\begin{equation}
    Om(z)=
    \frac{\left[H(z)/H_0\right]^2-1}{(1+z)^3-1}\, ,
    \label{eq:Om_statistic}
\end{equation}
and shown in Fig.~\ref{fig:Om_statistic}. By construction, for a flat $\Lambda$CDM model this quantity is constant and equal to the present matter density parameter, $Om(z)=\Omega_m$. Therefore, any redshift dependence of $Om(z)$ provides a useful diagnostic of deviations from the $\Lambda$CDM background expansion.

For the GA-reconstructed $F(R)$ model, we find only a very small departure from the constant $\Lambda$CDM behaviour. As shown in the inset of Fig.~\ref{fig:Om_statistic}, the reconstructed curve lies slightly above the $\Lambda$CDM value at low redshift and slowly approaches it as $z$ increases. This is consistent with the previous results: the GA reconstruction produces only a mild late-time modification, while the background evolution rapidly becomes indistinguishable from $\Lambda$CDM at higher redshift, where the imposed matter-era matching condition forces the model to recover the standard behaviour.

For comparison, we also show the corresponding $Om(z)$ obtained by fitting the same dataset within the CPL parametrization, with best-fit values $\Omega_m=0.303$, $H_0=66.9\;{\rm km\,s^{-1}\,Mpc^{-1}}$, $w_0=-0.888$, and $w_a=-0.22$. In this case, the deviation from a constant $Om(z)$ is much more pronounced, especially at low redshift. This illustrates that, although both the GA and CPL models can introduce deviations from $\Lambda$CDM, the reconstructed $F(R)$ model is much more tightly constrained around the standard background evolution. The $Om(z)$ diagnostic therefore confirms that the GA best-fit behaves as a small HS-like deformation of $\Lambda$CDM rather than as a strongly evolving DE model.

\section{Conclusion} 
\label{sec:conclusion}
In this work, we explored small deviations around the Hu--Sawicki (HS) class of $F(R)$ gravity models using current background cosmological data, namely DESI DR2 BAO measurements and the Pantheon+ SNe Ia catalogue, as discussed in Section~\ref{sec:data}. To this aim, we developed a pipeline that interfaces a Genetic Algorithm (GA) reconstruction with the cosmological likelihood evaluation. In particular, the candidate $F(R)$ functions are generated using the \texttt{GATO} code, while the corresponding background expansion history is reconstructed and tested against the data with the \texttt{CANDI} cosmological code, as described in Section~\ref{sec:methodology}.

The theoretical viability of the GA-generated functions was tested by applying the stability and consistency conditions discussed in Section~\ref{sec:theory_checks}. The best-fit reconstructed model satisfies the required conditions over the curvature range considered, with $\mu^2_{\rm eff}>0$, $f_{RR}>0$, and $|f_R-1|\ll1$. In particular, the order of magnitude of $|f_R-1|$, is very close to the GR limit, in agreement with the expectations for viable HS-like models.

Our results show that current background data allow only mild deviations from $\Lambda$CDM. The best-fit GA reconstruction remains very close to the constant contribution $F(R)=-2\Lambda$ over the full redshift range considered. The deviation is largest in the low-curvature, late-time regime, where modified-gravity effects can in principle contribute to the accelerated expansion, while it is rapidly suppressed at higher redshift. This behaviour is consistent with the HS reference model with $n=2$ and $b=0.1$, and with the matching condition imposed in the matter-dominated regime, which forces the reconstructed model to recover the $\Lambda$CDM background at high redshift.

The comparison of the minimum $\chi^2$ values also supports this interpretation. The GA reconstruction provides a slightly smaller $\chi^2$ than $\Lambda$CDM, while the HS reference model gives a slightly larger value. However, the differences are of order unity and therefore do not indicate a statistically significant preference for one model over the others. Similarly, the $Om(z)$ diagnostic shows only a very small departure from the constant $\Lambda$CDM behaviour, in contrast with more flexible phenomenological parametrizations such as CPL, where the deviation is much more pronounced. 

Regarding the correspondence between $f(R)$ gravity and an effective DE evolution, the reconstructed effective EoS parameter $w_{\rm eff}$ shows only a tiny dynamical variation, with a mild transition from an effective quintessence-like regime to an effective phantom-like regime around $z \sim 1$. Overall, the reconstructed $F(R)$ model should therefore be interpreted as a small HS-like deformation of $\Lambda$CDM rather than as a strongly distinct cosmological scenario.

We did not perform model selection because the effective number of parameters in the GA reconstruction is not well defined, so standard information criteria cannot be applied. Therefore, the comparison of $\chi^2$ values should be interpreted solely as a goodness-of-fit diagnostic.

A limitation of the present analysis is the computational cost of the full reconstruction pipeline. For this reason, we restricted the GA prior to smooth and small deviations around the HS model, rather than allowing completely arbitrary symbolic functions. While this choice is physically motivated and ensures that the reconstructed functions remain close to a viable region of theory space, it also limits the class of models explored. Future work should therefore extend the analysis to broader functional priors, different HS parameter choices, and more general $F(R)$ structures, in order to assess whether current and upcoming data show any preference for departures from the HS-like branch considered here. It will also be important to include perturbation-level observables, such as growth-rate and weak-lensing data, where modified gravity effects can be more directly constrained than with background distances alone.

A further natural extension of this work would be to perform a direct comparison with phenomenological DE parametrizations, in particular the CPL model. This comparison would be useful, since CPL is a common reference model for studying deviations from $\Lambda$CDM at the background level. However, it is not straightforward within the present version of the pipeline. Indeed, our reconstruction starts from a given $F(R)$ function and then computes the corresponding background expansion, while an $F(R)$ model exactly reproducing a CPL expansion history is not available in the literature. Therefore, a consistent comparison would require either extending the pipeline to include phenomenological $H(z)$ parametrizations under the same likelihood and numerical assumptions, or reconstructing the corresponding $F(R)$ representation of CPL. 

%\newpage

\acknowledgments
CDL and EF thank the Instituto de F\'isica Te\'orica, where part of this work was carried out, for warm hospitality.
CDL thanks financial support from the research grant number 2022E2J4RK ``PANTHEON: Perspectives in Astroparticle and Neutrino THEory with Old and New messengers" under the program PRIN 2022 funded by the Italian Ministero dell’Universit\`a e della Ricerca (MUR).
The work of EF is partially supported by the research grant number RM123188F70EF8AC ``Interstellar dust polarization as a CMB foreground - preparation of the LSPE measurements" under the Progetti di Ricerca (Piccoli, Medi) - Progetti Medi of Sapienza University of Rome. 
CDL and MM acknowledge financial support from Sapienza Università di Roma, provided through Progetti Medi 2021 (Grant No. RM12117A51D5269B). This work made use of Melodie, a computing infrastructure funded by the same project, and PLEIADI, a computing infrastructure installed and managed by INAF. SN acknowledges support from the research project PID2024-159420NB-C43, and the Spanish Research Agency (Agencia Estatal de Investigaci\'on) through the Grant IFT Centro de Excelencia Severo Ochoa No CEX2020-001007-S, funded by MCIN/AEI/10.13039/501100011033.

\bibliographystyle{apsrev4-2}
\bibliography{apssamp}

\end{document}